# Filamentary flow of vortices in a $Bi_2Sr_2Ca_1Cu_2O_{8+\delta}$ single crystal

L. Ammor[1*], A. Ruyter[1], V. A. Shaidiuk[1], N. H. Hong[2], and D. Plessis[3]

1. Université de Tours - LEMA- CNRS - CEA UMR 6157 - Parc de Grandmont - 37200 TOURS - France,

2. Department of Physics and Astronomy, Seoul National University, Seoul 151-742, Korea

3. CEA/DMAT - B.P. 16 - Le Ripault - 37260 MONTS - France,

Using micro-bridge technique, we have studied the vortex dynamics in a very low temperature region (*i.e.* $T/T_C \rightarrow 0$) of the B-T phase diagram of $Bi_2Sr_2Ca_1Cu_2O_{8+\delta}$ single crystal. We distinguish two types of vortex dynamics near the depinning threshold depending on the magnitude of the vortex-vortex interactions. For $0.01 \leq \mu_0 H < 1T$, we show that current-voltage characteristics (I-V) are strongly dependent on the history of magnetic field and current cycling. The sharp peak, so called "peak effect" (PE), observed in $\mu_0 H-I_c$ curve is due to a metastable state that can be removed after current cycling. At low field, I-V curves show steps that would be clearly related to "fingerprint phenomenon" since the relationship $R_d = dV/dI$ exists. This can be attributed to vortices flow through uncorrelated channels for the highly defective lattice. Indeed, as field sufficiently increases, these peaks merge to make broader ones indicating a crossover from filamentaty strings to braid river like in which vortex-vortex interactions becomes significant. As confirmed by the discontinuity in the critical exponent value $\beta$ determined in the vicinity of the threshold current using the power-law scaling $V \sim (I-I_c)^\beta$ with a crossover from $\beta = 2.2$ to $\beta = 1.2$. The strong vortex correlation along the c-axis has been clearly demonstrated using the dc-flux-transformer geometry for transport measurements that confirms the pseudo-2D behaviour of the FLL. Our transport studies are in good agreement with simulations results of 2D elastic objects driven by repulsive interactions through a random pinning potential.

## I. INTRODUCTION

Phase transition in the flux-line lattice (FLL) has been a subject of much interest from both theoretical and experimental points of view. In weak-pinning type-II superconductors, various topological phase transitions from a quasi-ordered FLL (elastic solid or Bragg-glass) to a flux-line liquid, or from a quasi-ordered FLL to a disordered FLL (plastic solid or vortex-glass) were predicted and observed experimentally [1-3]. Peak effect (PE), i.e. a sharp increase in the critical current $\mu_0H$-$I_c$ curve, is an important observation for tracking these topological phase transitions in both low- and high-$T_c$ superconductors [4-7]. In the low-Tc superconductors, the PE appears near $\mu_0H_{c2}$, when a transition from ordered to a strongly disordered pinning state occurs in the vortex lattices [8,9]. In high-$T_c$ superconductors, a pronounced PE or fishtail, in which Jc shows a sharp increase, was also observed and attributed to a transition from order to disorder (as in [10,11]), or from three dimensional (3D) to 2D phase (as in Refs [12-13]). It has been shown that there must be a big similarity in such metastability and hysteresis of I-V curves concerning these two peaks in both low- and high $T_c$ superconductors [14,15]. The question is whether the origin and/or the nature of these two phenomena would be the same or not.

The origin and the nature of the PE phenomena are still controversial issues. Since the first explanation of this effect suggested by Pippard [16] and the formal treatment made by Larkin and Ovchinnikov [17], a diversity of theoretical pictures has been proposed [see Ref.18]. Experimentally, the PE has been extensively studied, especially in the low-$T_c$ 2H-NbSe$_2$ single crystals [19]. In those studies, the PE was attributed to a structural transition from elastic vortex lattice (ordered) to a plastically deformed (disordered) vortex structure. On the other hand, in numerical studies, this peak is usually ascribed to plastic vortex depinning that was followed, at high velocity, by dynamic ordering of the vortex lattice [20,21]. Reports on Fe-doped 2H-NbSe$_2$ single crystals have given another explanation for the PE phenomenon [22] by suggesting that an edge contamination mechanism in the PE regime should activate the defective plastic flow within the bulk's part of the sample. The metastable and disordered phase appears at the sample edges and then dynamically coexists with the quasi-ordered phase in the volume of the sample [23]. Magneto-optical imaging of the vortex structure in high-$T_c$ Bi$_2$Sr$_2$CaCuO$_{8+\delta}$ (BSCCO) has shown similar results [24]. However, some work on amorphous Mo$_x$Ge$_{1-x}$ films and BSCCO single crystals with different contact geometries showed contrary results [25,26] by demonstrating that the edge effect is not important for static or dynamics vortex properties. Since there are so many different assumptions for the mechanism of this phenomenon, we aim to find out a possible way to clarify its origin.

PE and history effects have been widely studied by transports experiments in the low-$T_c$ materials. Measurements of I-V characteristics vs temperature (T) and magnetic field (B) have only been used to investigate the nature of pinning and structural transitions in vortex state in the low-$T_c$ materials [4,9,19,23]. However, in high-Tc superconductors, those measurements were technically impossible to perform, due to the problem of contacts heating and the limitation of the applied current. In order to avoid that, one must apply only a very small current. However, in this case, it is hard to detect signals. To overcome such a big difficulty, we have used a microbridge technique to be able to measure I-V characteristics at very low temperatures (*i.e.* $T/T_C \to 0$). Then, we can focus on the pinning-depinning mechanism and the dynamics of vortices in $Bi_2Sr_2Ca_1C_2O_{8+\delta}$ (Bi2212) single crystals deep inside the domain of irreversibility. In one of our previous work [Ref.27], we presented one possible way of interpretation. We proposed that the pinning by the surface disorder can not be ignored for understanding the vortex lattice dynamic in Bi-2212. In this paper, from the detailed observations, differential resistance curves and power-law scaling, our result implies a dynamic nature of the PE.

## II. EXPERIMENT

Samples used in this study are slightly over-doped Pb-doped Bi-2212 single crystals ($Bi_{1.8}P_{0.2}Sr_2CaCu_2O_{8+\delta}$) with a typical in-plane penetration length $\lambda_{ab} \approx 1700$Å. They were grown by a self-flux technique as described elsewhere [28]. A single crystal was extracted by cleavage, and then was shaped as thick micro-bridges (with a controlled size of 200 x 400 x 100 $\mu m^3$) by using our laser method [29, 30]. The advantage of this technique is to make it possible to reach low temperatures and to apply large current densities with a good homogeneity. The resulting sample was post-annealed by using an appropriate chemical treatment under an oxygen gas flow. Electrical contacts ($\leq 1\Omega$) were made by bonding gold wires by silver paste. The zero-field Ohmic resistance as a function of the temperature did not exhibit any anomaly. The crystal used in the experiment have a superconducting transition temperature $T_c=79.5$K with a transition width $\Delta T_c$ of about 0.5K, which confirms that the sample have a good homogeneity. The DC transport measurements were performed using a standard dc-flux-transformer geometry. The current is applied perpendicular to a magnetic field direction. I-V characteristics were obtained with a voltage resolution of 1 nV and a temperature stability better than 5 mK.

## III. RESULTS AND DISCUSSION

In order to investigate the influence of the history of current, field, and temperature cycling on the pinning of the flux-line lattice, we have performed current-voltage measurements at a fixed T = 5 K (*i. e.* $T/T_C \rightarrow 0$) and an applied magnetic field that is parallel to the c axis, ranging from 0.01 to 9T. Typical results are summarized in Fig. 1, where two distinct regimes could be seen. In Fig. 1, the inset shows the typical V(I) curve obtained for high magnetic fields ($1T \leq \mu_0 H \leq 9T$) while the main figure is for low fields ($0.01T \leq \mu_0 H < 1T$). Additionally, we have compared the effect of field-cooling (FC) under different cooling rates and of a Zero Field Cooling (ZFC). We have measured the same dissipation in the time-scale of our experiment. In particular, no effect is observed on the critical current in both ways either when the applied current is increased or decreased. One can see that when the magnetic field is decreased, different behaviours were observed in a restricted region of the phase diagram. Since the vortex lattice could appear at a low magnetic field, the I-V curves exhibit an S-shape with a high threshold current I*, but only for the first increase of the current. After this initial ramp, a reproducible $I_c < I^*$ can always be detected. Below 0.01T, we do not observe any hysteresis in the I-V curves. The two thresholds, $I_c$ and I*, identify two distinct states of the FLL, with one is more strongly pinned than the other. We have observed a hierarchy in the accessible threshold currents. Fig. 2 shows the field dependence of both I* and $I_c$ as a function of magnetic field. Each point, in I*($\mu_0 H$) curve, represents the threshold current measured in the first ramp up. In contrast with the $I_c(\mu_0 H)$ curve that decreases monotonously for magnetic field higher than $\mu_0 H = 0.07T$ (*i.e.* $a_0 \approx 1700 Å \approx \lambda_{ab}$), we note that I* curve has a pronounced peak at $\mu_0 H = 0.1T$ and it decreases rapidly as the applied field increases. This phenomenon is known as the "peak effect", which in most weak-pinning superconductors, is usually associated with a transition from elastic to plastic depinning [31].

A large value of the threshold current I*, obtained only for the initial ramp up of the current, was previously observed in pulsed-current experiments on $Bi_2Sr_2CaCu_2O_8$ single crystals, and this high threshold current state was enabled to a metastable state with a very long relaxation time [14,32]. A similarly strong metastability and history dependence were observed in both DC and pulsed transport studies of flux line lattices in the low-$T_c$ superconductor 2H-NbSe$_2$ ($T_c$=6.1K) [33,34]. In the PE regime, the first current ramp shows a large hysteretic critical current for the FC 2H-NbSe$_2$ sample, suggesting that vortices are strongly pinned. Once the flux lines are depinned, and have a subsequent ramp, the critical current takes a smaller value. More, the DC and AC magnetization studies in CeRu$_2$ ($T_c$=6.3K) and 2H-NbSe$_2$ also show a presence of a highly pinned and disordered vortex state when the sample is prepared using a FC process [35,36]. Ravikumar *et al.* also found the presence of a highly disordered vortex state when a sample is cooled under in $\mu_0H<\mu_0H_p$ by the DC magnetization technique [37]. This metastable FC state can be driven to the stable ordered state if one uses a slow ramp rate for the applied current [38]. The metastability of the FC state, near the peak effect, has been directly revealed in the mixed state of 2H-NbSe$_2$, Nb and CeRu$_2$ by using small angle neutron scattering combined with *in situ* transport and magnetic susceptibility measurements [39-41]. In high-Tc superconductors, such as BSSCO [13], BaKBiO$_3$ [42] and La$_{1.9}$Sr$_{0.1}$CuO$_4$ [43], several groups observed that Bragg peaks associated with FLL order were present for small magnetic field below (~0.05T for BSCCO) but disappeared quickly when this field was increased confirming a diordered FLL at high field.

Another remarkable behaviour in Fig. 1 is the multiple steps in the I-V curves measured upon increasing or decreasing bias current near the PE. For fields larger than 1T (*i.e.* a$_0 \approx$ 450Å $\approx \lambda_{ab}/3$), we do not observe any "step" in the I-V curves. In order to see the above features more clearly, we present differential resistance curves R$_d$=(dV/dI) at fixed magnetic fields in Figs. 3 and 4. The differential resistance R$_d$ is calculated by taking numerical derivatives of the curves that were shown in Fig. 1. From Fig. 3, one can see the current dependence of R$_d$ at 0.5T as a typical example. R$_d$ shows jagged peaks as "fingerprints" at the onset of motion. The fingerprints were observed with I both increasing (first ramp) and decreasing (note that the same feature was obtained for 0.07 T $\leq \mu_0$H < 1 T). The Fig. 4 shows that the jaggedness in R$_d$ disappears at higher field such as $\mu_0$H = 3, 5 and 9T. For fields larger than 1T, R$_d$ shows a pronounced maximum, which rapidly decreases to a terminal asymptotic value at high currents. The location of this peak **shifts** to lower current as the magnetic field increases. This behaviour is in contrast with that in elastic flow region where R$_d$ increases monotonically from zero and saturates at larger I [9].

The comparison with numerical simulations studies suggests that the peak in $R_d$ signifies a plastic flow region where the vortex matter moves incoherently and a coherent flow is recovered at high current [44,45]. This confirms that interactions are more important despite a pure elastic depinning process that has not been reached. Indeed, for elastic flow, $R_d$ increases monotonically from zero and saturates at larger I to the terminal value where the coherent flux flow is recovered. Here, the observed vortex dynamics confirms that the vortex flow morphologically changes from filamentary strings to a braided river as $\mu_0 H$ is increased, which has been proposed in recent numerical simulations results of superconducting vortices driven by repulsive interactions through a random pinning potential [46,47].

This plastic flow regime is also accompanied by a pronounced "fingerprint" effect. The jaggedness or "fingerprints" are generally attributed to the opening and the closing of channels as the applied current increases. Our data strongly suggest that the onset of flux motion in the region of the PE is in the form of defective and plastic flow, in which some parts of the FLL move, while others remain pinned. This type of dynamics was already observed in the three-dimensional 2H-NbSe$_2$ in the field range of plastic flow with filamentary vortex motion, and was interpreted as evidence of defective flow [9]. Such behaviour was also observed at sufficiently low temperature in amorphous Mo$_{77}$Ge$_{23}$ superconducting thin films [48]. This result is consistent with previous numerical work of Gronbech-Jensen *et al.* showing that at zero temperature, some filamentary flow channels are stable in a finite range of bias current [49]. They found that the transition between different flow channels' structures as the driving force varies would cause steps in the I-V curves. A more striking result was observed in the behaviour of the flux flow noise near the PE [50,51]. The large flux flow noise, in the plastic flow regime of the dynamics, was ascribed to an incoherent flow of a defective moving phase, as well as to a coexistence of moving and pinned vortex phases. Additionally, it was suggested that an important distinction between plastic and elastic response could be found in the correlations of the time-averaged velocity length, which characterizes the partial inhomogeneity of a moving FLL [52]. Indeed, in a model where dislocations are forbidden and the response is thus elastic, the time-averaged velocity will be partially homogeneous and correlated over the whole system. In contrast, in a system that exhibits a plastic flow, the time-averaged velocity will be spatially inhomogeneous and it yields, at least, a bimodal structure of the corresponding histogram [49].

This regime has been referred to as a plastic flow and may be associated with the memory effect and even hysteresis in the I-V curves.

Now let us turn back to the vicinity of $I_C$. By focusing on the nonlinear dynamics above the onset of motion, Fisher theoretically predicted that elastic depinning would show criticality and that velocity vs force curves in the vicinity of a threshold force $F_c$ would scale as $v = (F-F_c)^\beta$ where $\beta$ is a critical exponent [53]. This scaling law has been studied extensively in 2D charge density waves (CDW) systems where $\beta = 2/3$ [54]. It is however, not known whether this exponent occurs in other systems undergoing elastic flow or not. Fig. 5 shows a typical logarithmic plot of these fits close to the critical current. The exponent $\beta$ was determined by using the linear part of the I-V curves at low $[(I-I_C)/I_C]$ values. When varying the value of $I_c$ around the previously determined values (*i.e.* in the confidence interval), the value of $\beta$ changes in a way that to fit the data near the onset of vortex motion and, thus, allows us to estimate the error of $\beta$ which is determined to be about $\pm 0.1$. We show different exponent respectively in the vicinity of the peak regime ($\mu_0 H = 0.05$ and 0.15T), and in high field ($\mu_0 H = 3$ and 5T) with respect to $\beta \sim 2.2 \pm 0.1$ and $\beta \sim 1.2 \pm 0.1$. One has to remark that the V(I) curves at the depinning onset are never observed convex. The exponent $\beta$ is never lower than 1. Fig. 6 shows the field dependence of the exponent $\beta$ measured in the vicinity and above the peak regime. Note that exponents in these two regimes represent different physical phenomena. One can notice that in the vicinity of $I_c$, there is a difference in the vortex dynamics depending of the strength of the vortex-vortex interactions. At low field, vortex starts to move through non-interacting channels. On the other hand, when interactions between vortices are strong enough, braided river of vortices can be depinned while others remain pinned. At this point, the correlation of the vorices has to be checked to verify the pseudo-2D behaviour of the FLL. Thus, transport measurements have been carried out in the so-called "top" and "cross" configurations, as a function of the temperature T with a current I = 10mA [55]. The directional feature becomes more pronounced as the temperature decreases (see Fig. 7). The strong vortex correlation along the c-axis has been clearly demonstrated in the dissipative regime by transport measurements using the dc-flux-transformer geometry. Thus, we can argue that vortices have an infinite tilt modulus $C_{44}$ that may suggest us to consider the FLL as a pseudo-2D array of elastic objects. These features of plastic flow are in good agreement with previous experimental studies on colloids in 2D driven by an electric field and interacting with a disordered substrate. In this study, where plastic depenning with filamentary or river like flow has been found, the velocity - force curve obeys the scaling law with $\beta$ = 2.2 [56].

Our important findings, i.e. the PE in the critical current and the evolution of the differential resistance with increasing $\mu_0 H$ remind us about the reports on a "quasi-two-dimensional (2D)" and a "three-dimensional (3D)" flux-line lattice in the layered low-$T_c$ superconductor $2H\text{-}NbSe_2$ [57,58] describing the nonlinear transport properties of the FLL in this system. They used magnetic field dependence of the critical current and differential resistance to investigate the crossover of the dynamics from an "elastic flow" regime to a "plastic flow" one. It is found that the dimensionality effects are more pronounced in the disorder-dominated regime where $\beta \approx 1.3$ and $1.8$ for 2D and 3D case, respectively. In this regime, where the disorder dominates, $\beta$ presents the non-uniform filamentary motion of a defective FLL and the growth of a tenuous structure of connected paths that is similar to percolation. The apparent exponent in the plastic flow is in agreement with the numerical simulations investigations of dynamics of the 2D vortex system in strongly disordered Josephson junction arrays ($\beta \approx 2.2$) [59]. Other simulation studies in the plastic flow regime found $\beta \sim 1.2$ for driven vortex lattices in presence of random disorder [60], $\beta = 2.0$ for electron flow in metallic dots [61] and $\beta = 1.94 \pm 0.03$ for the colloid [62]. In contrast, the vortex flow behavior in the interaction-dominated regime weakly depends on dimensionality, $\beta \approx 1.2$, and the exponent indicates the collective motion of the FFL.

The general features associated with the peak effect in Bi2212 bear a striking similarity to those exhibited by the low-$T_c$ layered compound 2H-NbSe$_2$, where a PE exists. However, there are still two discrepancies. The first is that the PE and the associated metastable effects usually appear close to $\mu_0 Hc_2$ in NbSe$_2$; while in our case, it is restricted to a low field value. The second is, in NbSe$_2$, the dramatic change in the I-V curves by varying magnetic field, which is attributed to a structural transformation of vortices. Below the PE, an ordered phase is formed by vortex interactions while above the PE, pinning induces a disorder-driven non-thermal phase transition to a disordered phase. In contrast, our result implies a dynamic nature of the PE, in agreement with recent numerical and experimental works [63-65].

**IV. CONCLUSION**

By using micro-bridge technique, we have been able to measure vortex dynamics at T = 5 K (*i.e.* $T/T_C \to 0$) in a $Bi_2Sr_2Ca_1Cu_2O_{8+\delta}$ single crystal. The current-voltage characteristics (I-V) are found to be strongly dependent on the current cycling. As consequences, the observed peak effect is attributed to a metastable state. However, a stable state can be obtained after a first ramp of applied current. A "fingerprint phenomenon", which was found in the current dependence of the differential resistance $R_d$ = dV/dI measured in the vicinity of the PE region (*i.e.* $1 \leq \lambda_{ab}/a_0 \leq 3$), strongly suggests that vortex flow is plastic and it occurs through filamentary channels for lattices with defects. This vortex depinning process changes when vortex-vortex interactions are strong enough. Then, the width of moving channels increases leading to braid rivers in which elastic interactions play a significant role. Additionally, we have shown that, at the threshold current, a scaling law exists between the current and the voltage both in the vicinity and above of the peak effect regime, $V \sim (I-I_c)^\beta$, with $\beta \approx 2.2$ and 1.2, respectively. All of these features are in good agreement with simulation results reported for 2D systems; in particular, for superconducting vortices driven by repulsive interactions through a random pinning potential. A strong vortex correlation along the c-axis was clearly seen while using the dc-flux-transformer geometry for transport measurements confirming the pseudo-2D behaviour of the FLL.

**Acknowledgments**

We gratefully acknowledge helpful discussions with A. Pautrat and Ch. Simon.

**Figure caption**

Figure 1: (a) Typical current-voltage curves after a FC measured at 5K. V(I) curve for $\mu_0H$ = 0.15 T: Arrows indicate increasing and decreasing applied current. The first increase of current defines I* (1: filled circle), the following decrease of the current defines $I_c$ (2: open square) and second increase of current (3: filled square). Note the jumps in the V(I) curve. Inset: reversible I-V curves measured at $\mu_0H$ = 5T after both increasing (filled circle) and decreasing currents (open circle).

Figure 2: Field dependence of both the threshold I* (filled square) and the critical current $I_c$ (open square) at 5K. Inset: microbridge (200*400*100$\mu m^3$) of Bi-2212 single crystal.

Figure 3: Variation of the differential resistance $R_d$ = dV/dI obtained from the I-V curves at 0.05, 0.15 and 0.5T showing a typical feature of "fingerprint" phenomena in the vicinity of the peak effect.

Figure 4: Evolution of the differential resistance $R_d$ with increasing $\mu_0H$ measured above the peak.

Figure 5: Typical scaling plots for the I-V characteristics at fields in the vicinity of the peak regime ($\mu_0H$ = 0.05 and 0.15T), and above the peak regime ($\mu_0H$ = 3 and 5T).

Figure 6: Field dependence of the exponent $\beta$ measured in the vicinity and above the hysteretic region. The exponents in these two regimes are extracted from I-V characteristics with first increasing (filled square) and decreasing (open square) current after field cooling.

Figure 7: $V_{Top}/V_{Bot}$ *vs.* T for I = 10mA at various fields 0.5 (filled circle), 1 (open circle), and 3 T (filled triangle). Inset : Cross view of the crystal along its length.

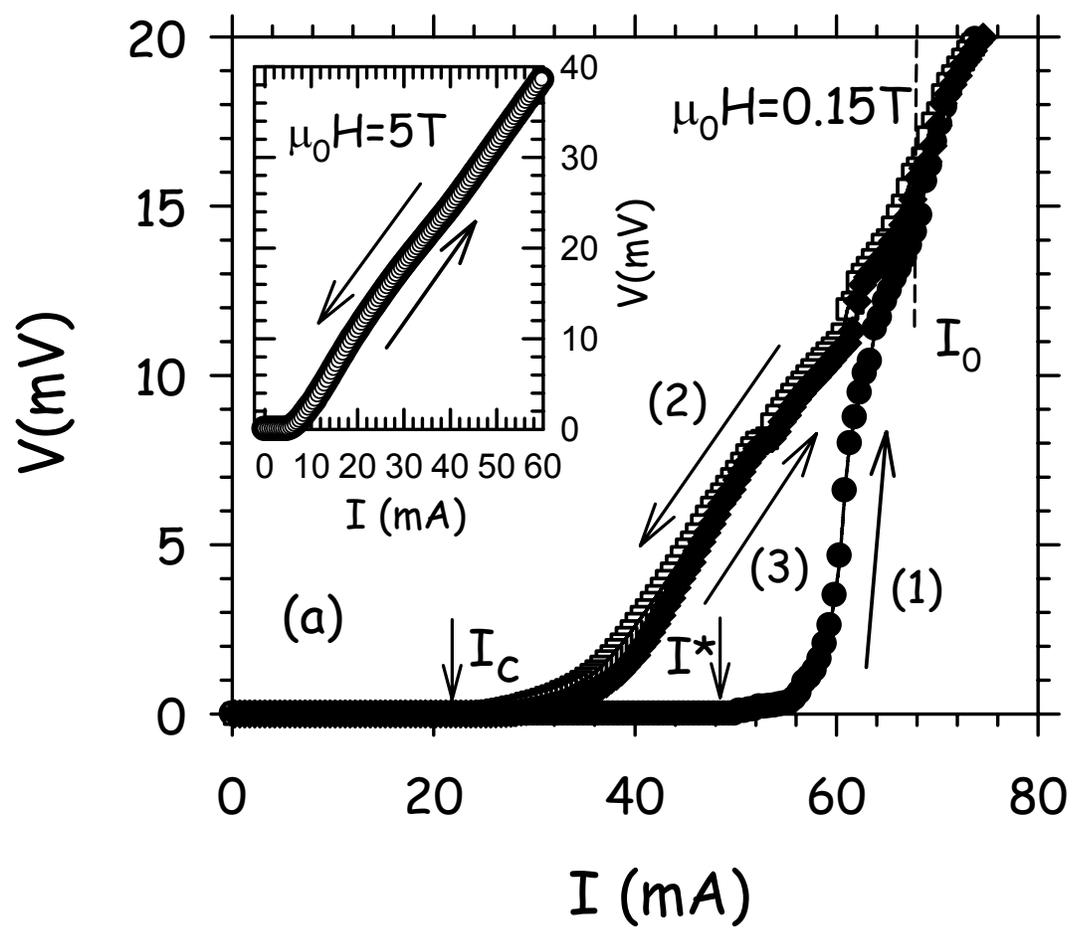

(a)

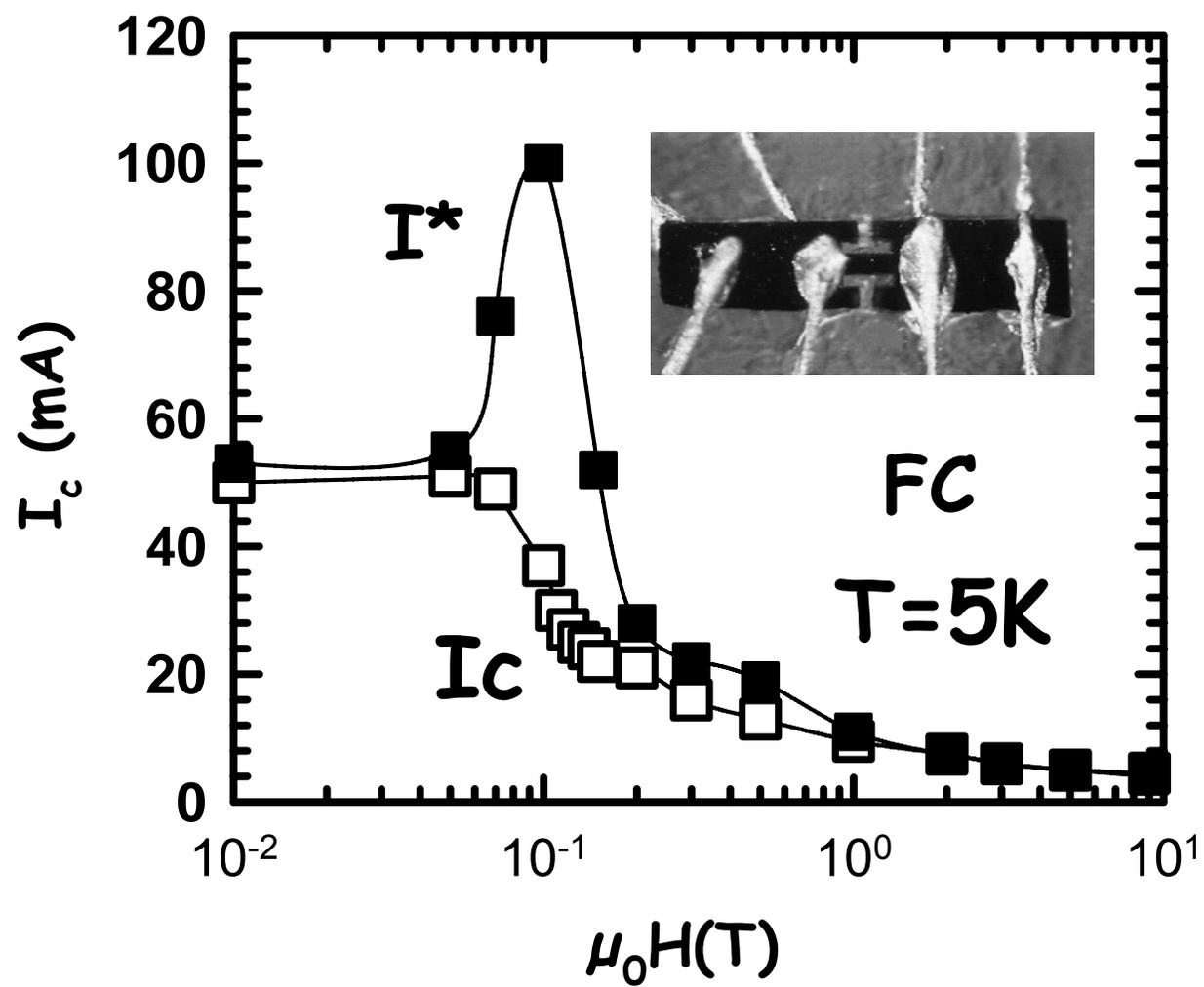

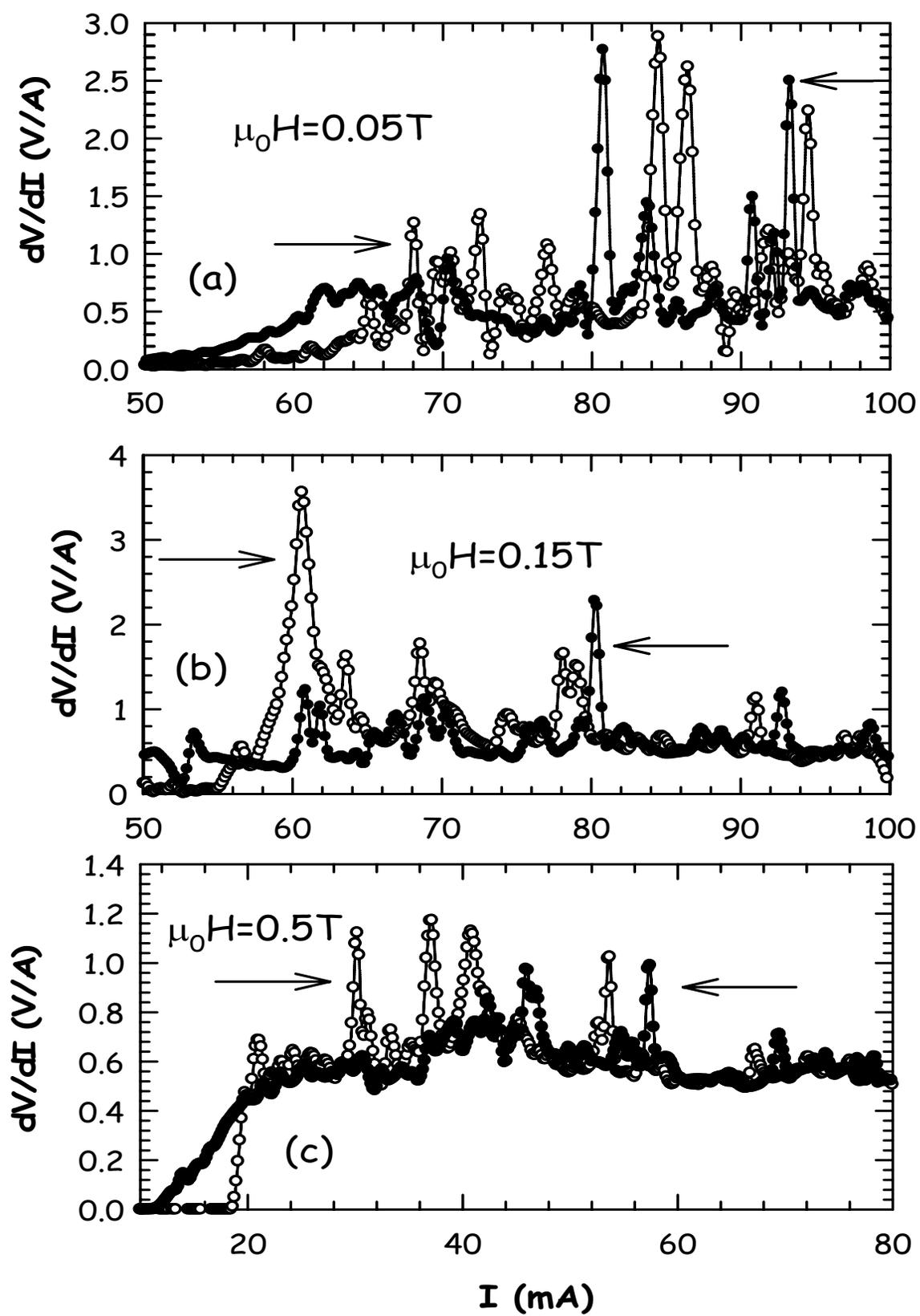

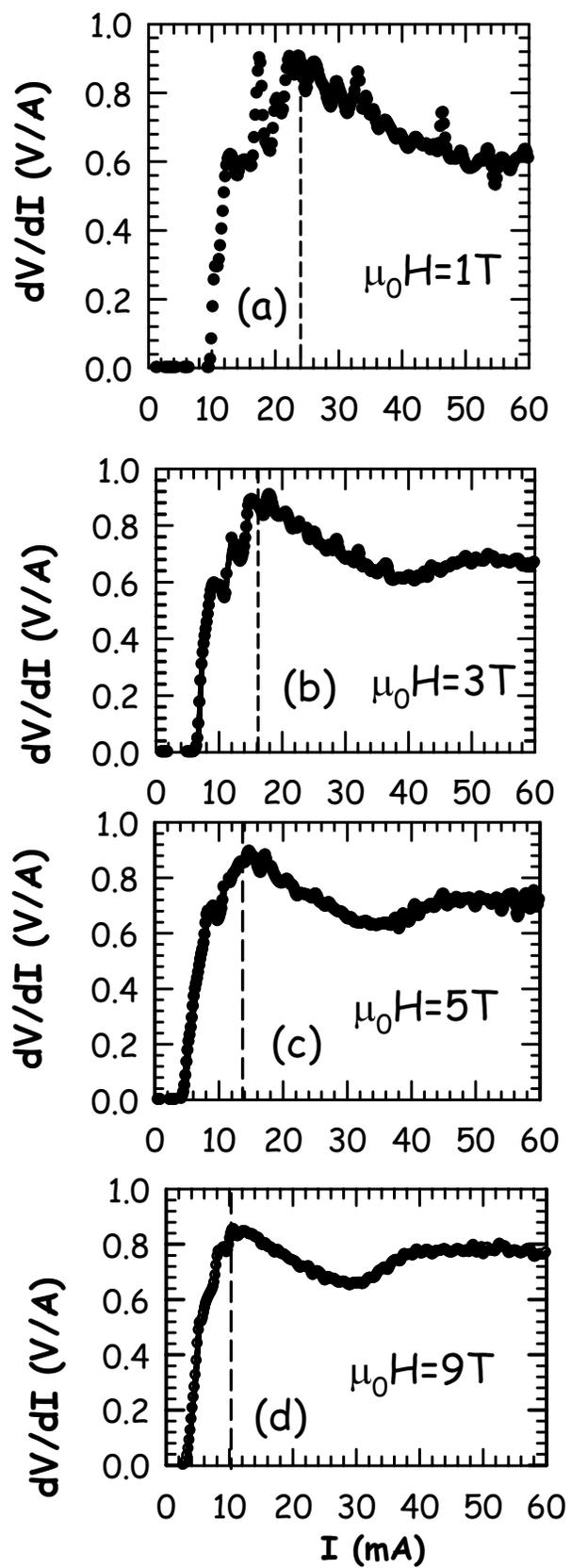

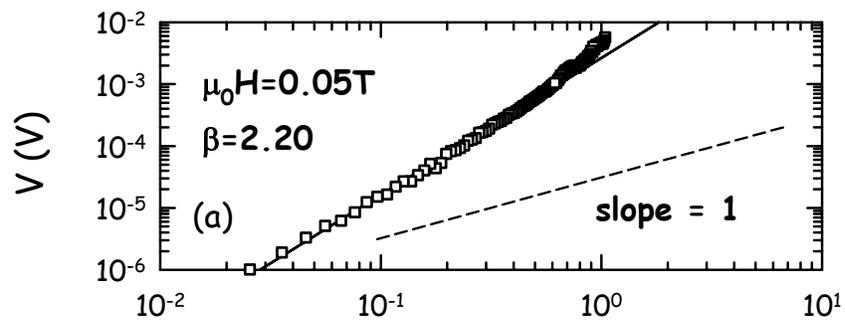

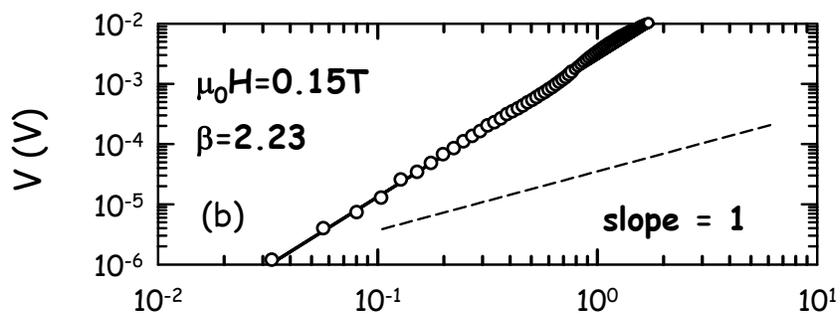

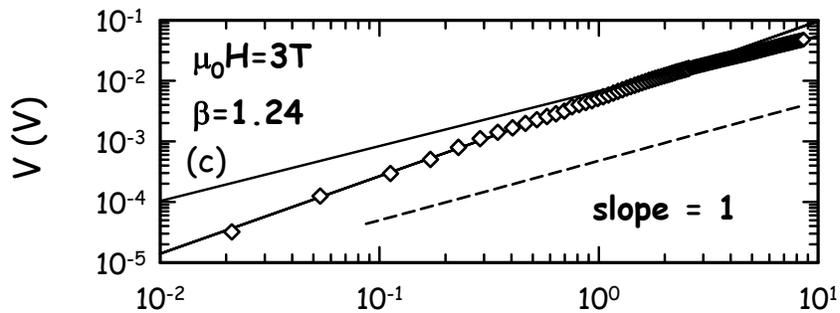

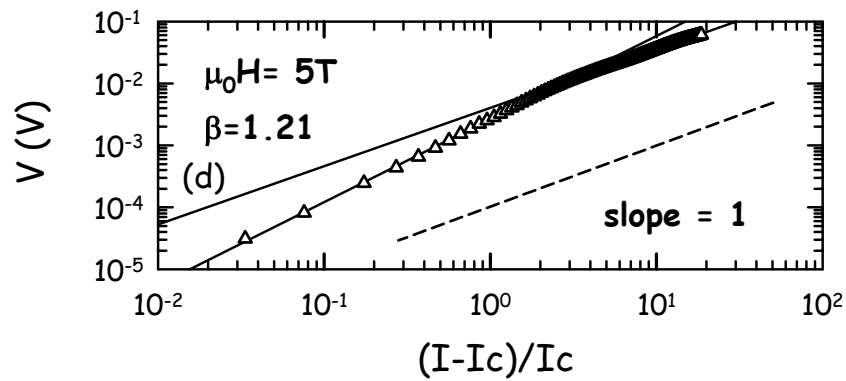

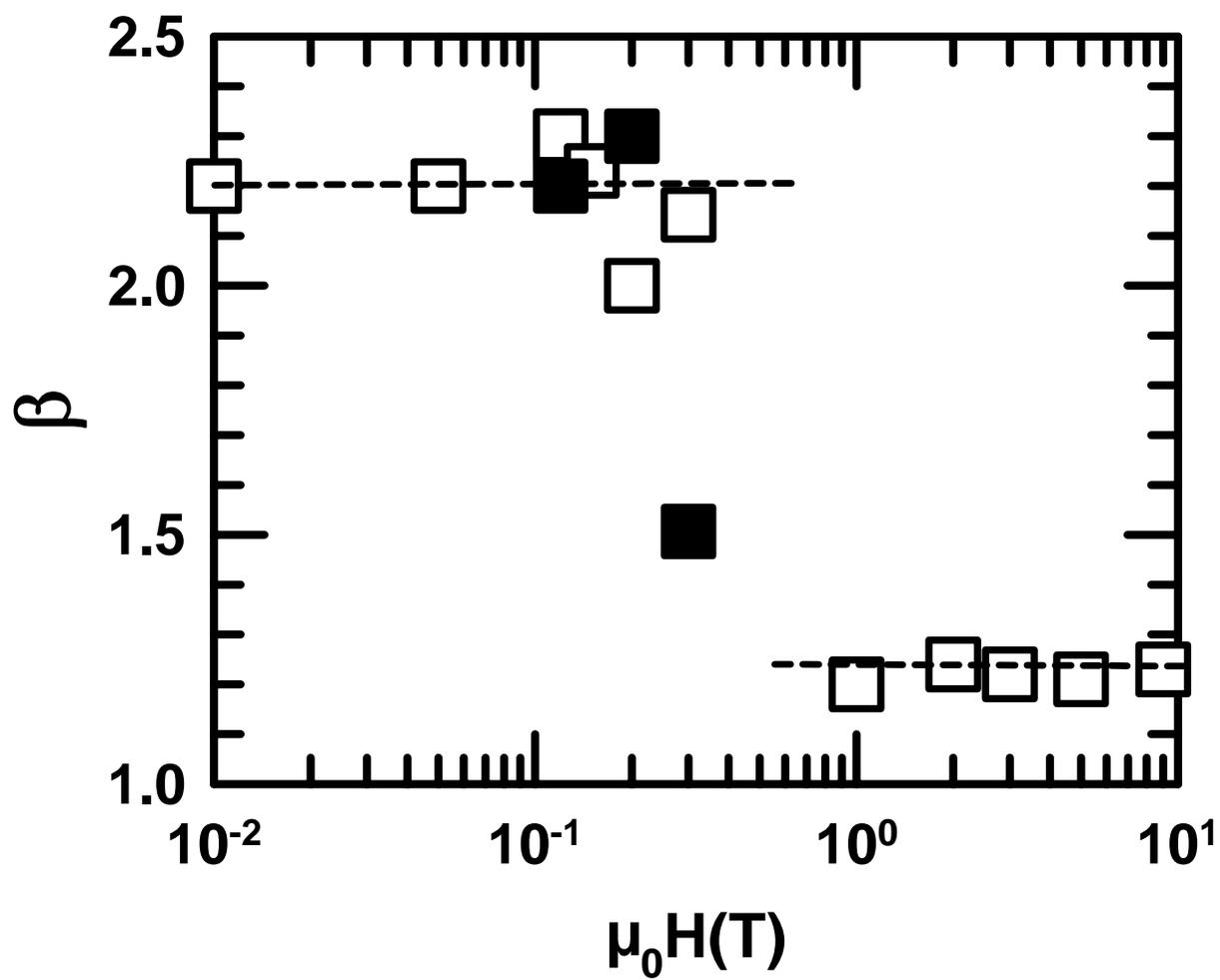

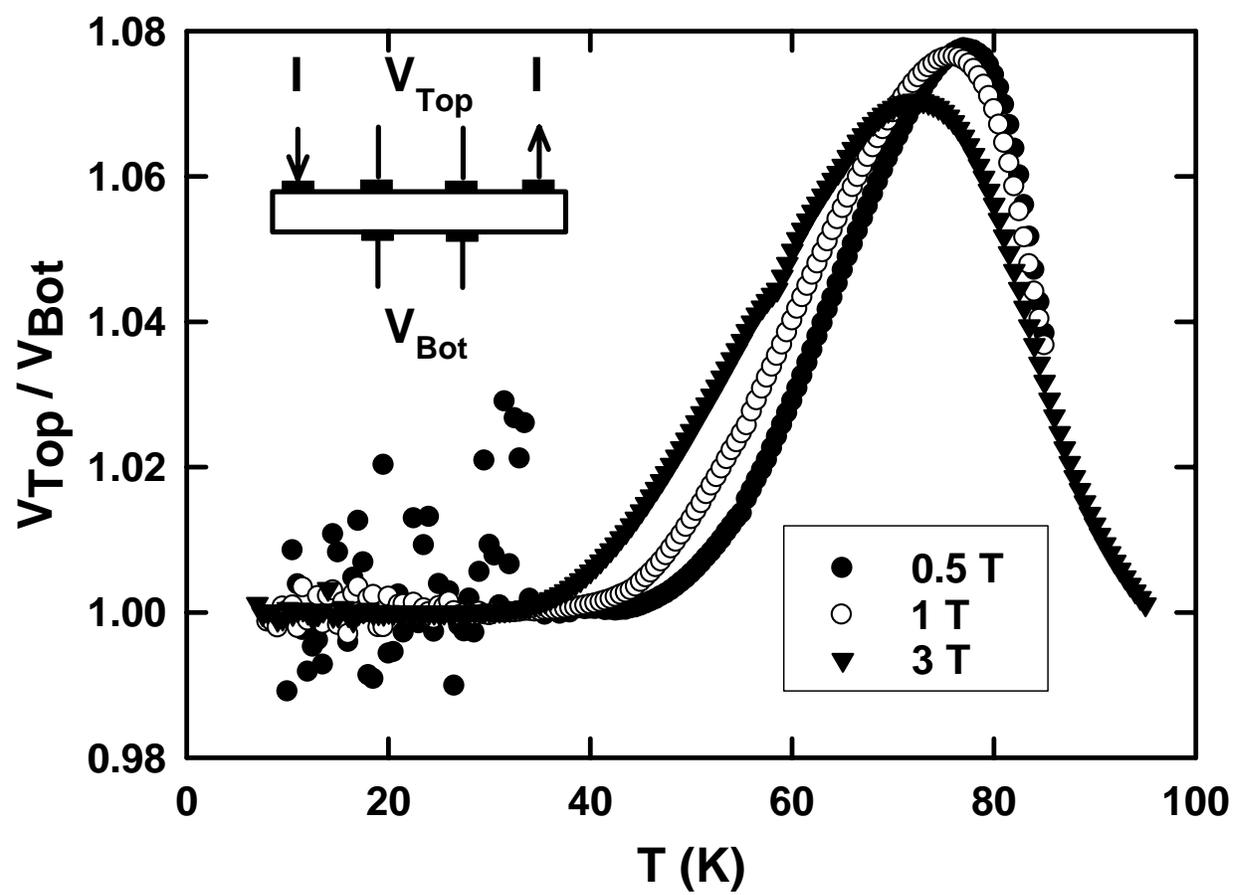